# By-Software Branch Prediction in Loops

Maziar Goudarzi, Reza Azimi, Julian Humecki, Faizaan Rehman, Richard Zhang, Chirag Sethi, Tanishq Bomman, Yuqi Yang

**Abstract**— Load-Dependent Branches (LDB) often do not exhibit regular patterns in their local or global history and thus are inherently hard to predict correctly by conventional branch predictors. We propose a software-to-hardware branch pre-resolution mechanism that allows software to pass branch outcomes to the processor frontend ahead of fetching the branch instruction. A compiler pass identifies the instruction chain leading to the branch (the branch *backslice*) and generates the pre-execute code that produces the branch outcomes ahead of the frontend observing them. The loop structure helps to unambiguously map the branch outcomes to their corresponding dynamic instances of the branch instruction. Our approach also allows for covering the loop iteration space selectively, with arbitrarily complex patterns. Our method for pre-execution enables important optimizations such as unrolling and vectorization, in order to substantially reduce the pre-execution overhead. Experimental results on select workloads from SPEC CPU 2017 and graph analytics workloads show up to 95% reduction of MPKI (21% on average), up to 39% speedup (7% on average), and up to 3x improvement on IPC (23% on average) compared to a core with TAGE-SC-L-64KB branch predictor.

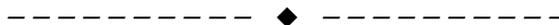

## 1 INTRODUCTION

Virtually all conventional branch predictors rely on the history of the branch outcomes. While effective in many workloads, such history-based predictors often fail on load-dependent branches (LDB) due to lack of predictable patterns in the branch history. Previous studies have shown the significance and impact of such LDBs [1], [2] in overall performance. Furthermore, LDBs incur long latency to resolve if the load misses in the caches, and thus, the pipeline is fed with many instructions from potentially incorrect execution path to be only squashed later on a mispredicted LDB.

Fig. 1 shows an example from Leela (SPEC 2017) with an LDB dependent on multi-level indirect loads. The key observations here are that (a) a loop repeats the LDB, and (b) the Taken/Not-taken outcomes of the LDB instances can be computed earlier in parallel since there is no loop-carried dependence. Existing hardware-only techniques [3] attempt to automatically form and pre-execute the branch *backslices* (i.e., the chain of instructions feeding the LDB). Major drawbacks of these approaches are: (i) cost, i.e., they need to dedicate hardware for backslice identification at runtime, (ii) latency, i.e., the initial iterations of the loop cannot benefit since hardware has not yet identified the pattern (this is especially a problem for loops with low trip counts, such as the one shown in Fig. 1), and (iii) aliasing, as multiple copies of the same branch (due to loop unrolling or other compiler optimizations based on the code strcuture) may confuse the hardware learning or engagement.

To address these challenges, we propose a SW/HW mechanism to pre-compute the branch outcomes in software and tell the BPU which dynamic instance of the branch they correspond to. This mechanism can be considered as the *branch-counterpart for software data prefetching*. If the branch information, which is computed in SW, is provided early enough to the HW (i.e., before the frontend observes that dynamic instance of the branch), a prospective misprediction is avoided. Even if late, the pre-computation reduces the branch misprediction penalty since the branch resolution latency is reduced because the data elements needed to resolve the LDB are brought to caches earlier, similar to [4]. We call this proposed mechanism BOSS, **B**ranch-**O**utcome **S**ide-channel **S**tream.

```
void FastBoard::kill_neighbours(…) {
    ...
    do {    ...
        for (int k = 0; k < 4; k++) {
            int ai = pos + m_dirs[k];

            if (m_square[ai] == kcolor) { // LDB
                ...
            }
        }
        pos = m_next[pos];
    } while (pos != vertex);
```

Fig. 1. An example LDB in a loop from leela, SPEC 2017.

We define the BOSS SW/HW interface by using memory-mapped I/O channels. As a result, we do not need any extension to the base processor ISA (ARM in our current evaluation). The memory layout for channels is defined such that multiple branch outcomes can be communicated to HW in parallel, thus enabling the vectorization of the pre-execute loop. BOSS needs two new operations: (i) **BOSS_open** operation for each BOSS channel to tell the Branch Prediction Unit (BPU) which static branch(es) this channel will be feeding, and (ii) **BOSS_write** operation to pass branch outcomes and their corresponding indexes to the BPU through the configured BOSS channel. Fig. 2 shows the instrumented source code for LDB in Fig. 1. Before the outer loop, the BOSS channel is configured by the `BOSS_open()` operation; then the pre-execute loop produces the branch outcomes and passes them to hardware via `BOSS_write()` operation, detailed below. Note that the pre-execute loop does not have to cover the *entire* iteration

---

● *Authors are with Toronto Heterogeneous Compilers Lab., Huawei Research Center, Toronto, ON, Canada. E-mail: maziar.goudarzi@, reza.azimi1@huawei.com.*

space of the target loop; the programmer (or compiler via Profile-Guided Optimization) may decide to only partially cover the iteration space.

```
void FastBoard::kill_neighbours(…) {
  ...
  BOSS_open(ch1, BR, End);
  do {
    // pre-exec loop
    for (int k = 0; k < 4; k++) {
      int ai = pos + m_dirs[k];
      bool outcome = (m_square[ai] == kcolor);
      BOSS_write(ch1, k, outcome );
    }
    ...
    // Target loop
    for (int k = 0; k < 4; k++) {
      int ai = pos + m_dirs[k];
BR:   // The b.ne  instruction for below if-statement
      if (m_square[ai] == kcolor) { // LDB
        ...
      }
    }
End:
    pos = m_next[pos];
  } while (pos != vertex);
}
```

Fig. 2. BOSS pre-execute loop for target branch LDB in Fig. 1. Note that the BR label corresponds to the conditional branch instruction for the designated if-statement, and that the End label designates an instruction after the inner loop exits.

BOSS is designed for loops where each dynamic instance of the target branch can be identified by a loop iteration counter. For other repetitions, such as recursions, this scheme does not work outright. We employ strip mining to partition loops with large trip counts into multiple loops with small trip counts. Any duplication of the target branch does not hinder BOSS since same channel can feed multiple static branch instances. Success of the scheme depends on two factors: (i) the pre-execute loop is run early enough before the frontend reaches the branch, and (ii) as in software-prefetching, the overhead of additional instructions does not overweigh the gains; unrolling and vectorization can help to reduce the overhead. The pre-execute loop can also be skipped statically (decision made by profiling) or dynamically (only executed if the target branch is mis-predicted beyond a set limit).

Our contributions are:
- A software-hardware mechanism for by-software pre-resolution of hard-to-predict branches in loops. This resembles software prefetching, but for branch prediction.
- Design of the SW/HW interface to allow vectorization of the pre-execute loop to reduce the scheme overhead.
- Design of the microarchitecture, and modeling and evaluation of the scheme on Gem5.

## 2 BOSS FOR SOFTWARE BRANCH RESOLUTION

### 2.1 System Overview

Each BOSS channel is configured by software to feed one (or more) static branch instances, and stores the branch outcome (**T**aken/**N**ot **T**aken) for the dynamic instances of the target branch, identified by their corresponding *iteration* number in the loop. The hardware monitors fetch, commit, and squash of the target branches to keep track of appropriate iteration number to consume next. Each iteration of the outer loop marks one *generation* of the inner loop. BOSS keeps track of the generation number that is currently in the channel (by monitoring the End PC, which was passed during channel configuration) so as to avoid mis-consumption in case of early break from the inner loop. Note that the frontend may speculatively go to multiple generations ahead by an early break or an otherwise exit from the inner loop. Nevertheless, the outcomes for only one generation are kept in the BOSS channel (obviously this can be extended). When a BOSS_write for a new generation commits, which is always later than the commit of all instances of prior generations, any remaining items from the previous generation in the channel are discarded.

### 2.2 SW-HW Interface

BOSS SW/HW interface consists of two operations: 1. **BOSS_open** for configuration of the channel per target branch (only done once for each branch) and 2. **BOSS_write** to pass T/NT branch outcomes of the chosen iteration-space of the loop to the processor frontend. We avoided introducing new instructions to the base ISA, and instead used ordinary memory write instructions (ST instruction in the ARM ISA) to a dedicated address-range. We allocate 8-Bytes for configuration and 256-Bytes for outcomes for each channel. The 256-Bytes address-range corresponds to the iteration-number (the dynamic instance number) 0 to 255 of the target branch; higher iteration numbers rotate over the same address-range. The choice of an address-range, instead of a single byte address, to pass the outcomes to hardware allows to vectorize the pre-execute loop unless a loop-carried dependence prevents it.

BOSS extends the architecture state, but since the added state is only a set of hints to the BPU, it is safe to ignore to save/restore it upon context switch. Nevertheless, ordinary load instructions (LD in ARM ISA) from same dedicated address-range are employed to find out open BOSS channels and read out the BOSS state of the open ones. Employing lazy save/restore, similar to the case for floating-point registers, is another possible improvement.

### 2.3 Compiler Support

The target branch is either annotated by the programmer, or automatically identified by PGO analysis on representative profiles. Then a compiler transformation pass produces the pre-execute loop through the following steps:
1. Identify the loop induction variable (loop counter), and its start, end and increment variables/values.
2. Starting from the target branch, trace back the chain of instructions feeding the branch back to the loop induction variable.
3. Generate code for the pre-execute loop before the target loop, with same range and increment. Within the loop, compute the branch-outcome for a given value of the induction variable and use BOSS_write to communicate the outcome.
4. Employ strip-mining if the range of values of the induction variable is large (more than the BOSS channel capacity) or not known statically.

The branch-outcomes are only useful if passed to BPU before the frontend reaches that instance of the branch. Thus, the pre-execute loop should be put as early in the code as possible. The compiler checks data dependencies for the above, but usually limits itself to the function boundary. Cross-module analysis can be employed to go even beyond the function boundary.

## 2.4 Microarchitecture

The branch-outcomes are stored in a lookup-table addressed by a <*channel*#, *generation*#, *iteration*#> tuple (Fig. 3); an outcome bit and a valid bit are stored per entry. On the outcome-production side (i.e. the branch-outcome values coming from the software), the channel# and iteration# are already implied in the address employed in the **BOSS_write** operation. The generation# is incremented under the hood when the End instruction (which represents an arbitrary instruction right after the loop ends, and whose relative-PC was passed during **BOSS_open** operation—See Fig. 2) commits.

On the consumption side, (i.e. the branch-outcomes fed to the BPU) the fetch/squash/commit of the target branch as well as the Loop-End instruction are monitored. For the target branch, the fetch/squash do an increment/decrement of the consumer iteration# respectively, and a commit removes the corresponding branch-outcome from the BOSS outcomes storage if there. For the Loop-End instruction, the fetch/squash do an increment/decrement on the consumer generation#, and also push+reset/pop the current consumer-iteration# on the iter#-stack table in the micro-architecture; note how this stack is required to properly handle squash of the Loop-End instruction. The commit of the Loop-End instruction increments the producer-side generation#.

The outcomes coming from BOSS take priority over ordinary BPU predictions since the BOSS outcomes are absolutely correct ones. Thus if a branch instance hits the BOSS storage, the stored outcome is passed as the prediction. The multiplexer on the top-left of Fig. 3 does this.

As in Fig. 3, two tables store target branch PCs and End-instruction PCs, two tables keep track of proper iteration# (one level depth for the stack was enough in all our experiments), and two other tables store producer-side and consumer-side generation#. For a typical case of supporting 4 simultaneous BOSS channels, 256 iterations and 2 generations per channel, a tiny amount of 256B (BOSS outcomes: 4 channels x 256x2 bits per channel) + 64B (2 PC tables: 4 channels x 8B PC each) + 8B (2 iter# tables: 4 channels x 8b iter# each) + 1 B (2 gen# tables: 4 channels x 1 bit each) = 329B of storage is enough.

Note that the entire BOSS system can be turned off until a BOSS_open operation is executed. An additional BOSS_close operation can signal to turn it off again after the outer loop.

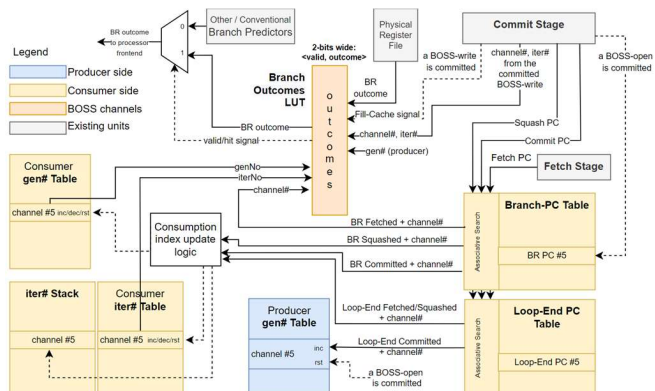

Fig. 3. Micro-architecture of the BOSS mechanism.

## 3 EXPERIMENTAL RESULTS

We implemented BOSS in gem5 and evaluated a number of applications from SPEC, GAP and cBench suites. Target branches and Regions-of-Interest (ROIs) were selected from prior studies [3], [5] as well as from our own perf analysis, and were evaluated on the processor settings in Table 1. Each ROI covers a continuous execution path such that covers several calls to the function enclosing the target branch. Note that for BOSS-instrumented cases, the pre-execute loop is also included in all the measurements.

TABLE 1: EXPERIMENTS SETUP

| Processor core | ARM v8-a, 8-issue, 192-entry ROB |
|---|---|
| Branch predictor | TAGE-SC-L, 64KB |
| L1 Instruction/ Data caches | 32KB-2way/64KB-4way, 64B lines, 2-cycle hit latency, write back |
| L2 unified cache | 2MB, 8-way, 20-cycle latency, write-back |
| HW prefetcher | Stride prefetcher for I/D L1/2 |

Fig. 4 shows the overhead instructions normalized to the baseline. Fig. 5 and Fig. 6 show speedup on each ROI, and the IPC gains respectively. For each experiment *loop*, *unrolled*, and *vectorized* bars correspond to applying no change, unrolling, and vectorization to the pre-execute loop. Although IPC is consistently improved significantly, the overhead decides whether that translates to speedup or not.

Experiments show that although unrolling and vectorization reduce the instruction overhead, but since they redue the number of instructions between the production and consumption sites as well, they are not always successful in getting better gains (see leela-kn for instance). Note that since this is a software-based optimization, it can be avoided upon slowdowns. Thus, we focus only on the sped up cases from now on.

The gains come from removing mispredictions from the target loop. Fig. 7 shows the obtained reductions in misprediction rates of the target branch. If applied early enough, BOSS should remove all mispredictions of the target branch, but since we did not go back beyond the function starting point to inject the pre-execute loop, this did not realize in many cases. Furthermore, when the trip-count of the loop is low, e.g. 4 and 8 in the case of leela functions, even the pre-execute loop instructions themselves are not numerous enough to provide the necessary gap between the commit of the BOSS-write instructions and the fetch of the target branch.

BOSS addresses specifically chosen branches. The reduction in total MPKI depends on the contribution of the chosen branch in total MPKI. Fig. 8 shows the MPKI of all branches and the obtained gains. Note that branches in the pre-execute loop are also counted in these results for *Loop*, *Unrolled* and *Vectorized* cases.

## 4 RELATED WORK

Various helper thread mechanisms [6], [7], [8] run a (potentially stripped down) thread ahead of the main thread, resolve the branches and inform the main thread. The obvious overhead and the synchronization difficulty between the two threads are the main concerns.

Runahead techniques [3], [9], [10] let the processor run original code earlier, or actively deduce loops and backslice of delinquent branches or the loads in them [13], and run them earlier, so as to resolve future branches. The learning latency, especially for low-tripcount loops, remains the limiting factor.

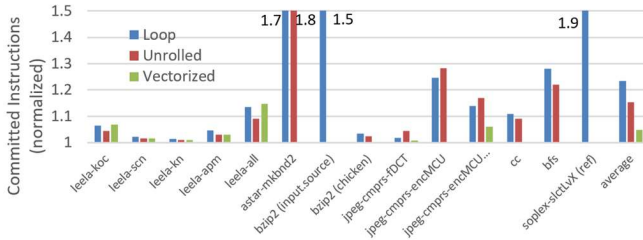

Fig. 4. Instruction overhead; the additional Committed Instructions.

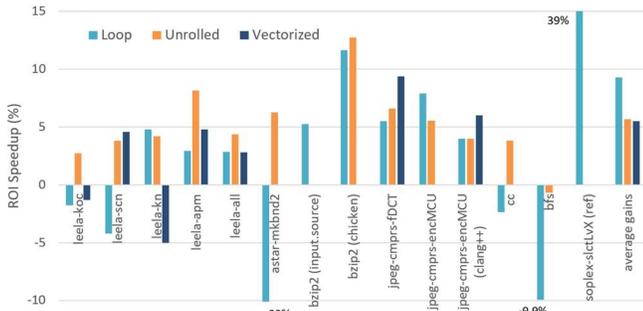

Fig. 5. ROI Speedup values.

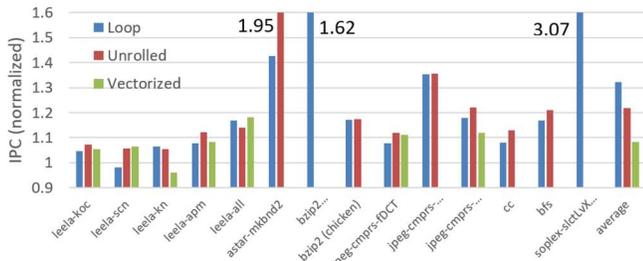

Fig. 6. IPC Gain on ROIs.

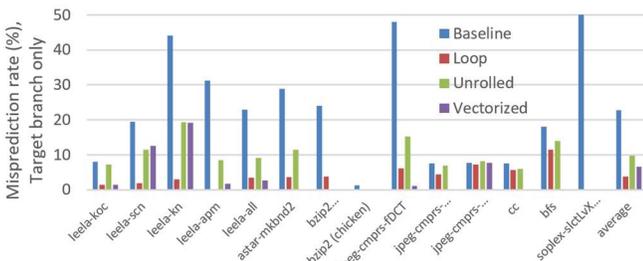

Fig. 7. Branch Misprediction rates on the Target Branch only.

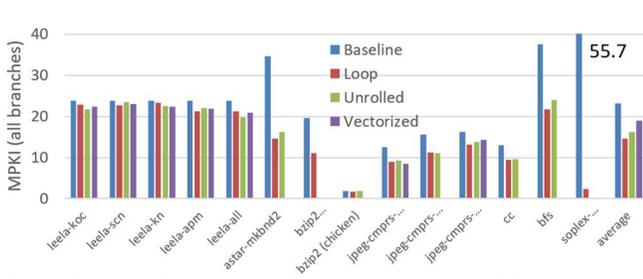

Fig. 8. Branch MPKI numbers for all branches in the ROI.

Customized branch predictors [2], [11] use either longer branch histories [2] or custom logic [11] for hard-to-predict branches. Longer histories do not help on load-dependent branches, and custom logic needs substantial change in the hardware. Correlating the branch with an earlier load-address or store instruction [12] is another customization mechanism. Complementarily, BOSS extends the HW predictor by SW.

### 4.1 Comparison with branch runahead

Branch Runahead [3] reports higher speedups, but note that it runs the additional instructions essentially on an extra, although stripped-down, core (so called DCE: Dependece Chain Engine), and thus the overheads are not reflected in the reported gains. BOSS, on the other hand, runs everything on the same core; consequently, BOSS does not require an additional core to work, and the gains reported here take the overhead into account as well.

As inherent in every full-hardware mechanism, there is a delay to detect the backslice instruction sequence with enough confidence; this is inherent because a full-hardware mechanism needs to see the same instruction sequence repeated at least a few times to build enough confidence level. For same reason, branch runahead is reporting that on average roughly 20% (up to 40%) of the time the mechanism is inactive (meaning that "at the time the core needed the prediction, no dependence chains had been activated to produce that prediction" [3]). A co-designed approach such as BOSS resolves this issue since the compiler has statically identified and formed the backslice; thus, no runtime overhead applies to detect it, nor it is missed by the detection mechanism.

Another problem inherent in all-hardware mechanisms is late initiation points; after the backslice is identified, the hardware mechanism needs to identify an initiation-point to copy/synchronize the input values of the backslice from architecture registers of the main core to the extra core, to let it start executing the backslices. Branch runahead does that upon detecting a branch misprediction on the registered branch; this is a good choice to get the latest live-in values for the backslice input registers, but may be too late to initiate execution of them. Branch Runahead [3] reports that on average another ~30% (up to over 60%) of the time the mechanism is late ("The late category refers to predictions which have active chains, but are generated too late to be useful for the core" [3]). In BOSS, since the compiler statically generates the backslices and has full visibility into all the codes around it, the pre-execute loop can be brought maximally forward to allow maximum available time between production and consumption of the branch outcomes so as to minimize the lateness.

A downside of a co-designed mechanism such as BOSS is that software/compiler does not know if the misprediction actually happens at runtime or not. Thus the overhead is always paid, whereas the in-hardwarem mechanism avoid it.

Branch runahead has in-hardware mechanisms to handle chained/nested branches. This is inexpensive in hardware because the branch-outcomes are nearby and can be used under the hood to trigger other backslices, but doing it in software requires adding new condition-checks in the pre-execute loop which may in turn get mispredicted. Thus, BOSS would not suit such cases of nested branches.

### 4.2 Comparison with control-flow decoupling

Control-flow decoupling (CFD) [5] changes the code struc-ture to produce branch outcomes earlier and explicitly consume them in a second loop.

While CFD has its advantage in avoiding a redundant compute of the branch condition, it is not practical in many cases for 3 important reasons:

1. CFD extends visible architecture state: the contents of the architectural branch-queues that hold the branch-direction outcomes; if these values are lost during context-switch, the program functionality fails. Same is true for function calls when the callee may use same CFD mechanism as the caller, and thus may overwrite those branch-direction outcomes. To address these issues, [5] provides Save_BQ/Restore_BQ interfaces, but this actually means that library-developers should also be aware of, and adhere to, this new architecture-state expansion, and similarly OS should also be aware of it and save/restore the newly added architecture-state beyond the ISA documented in the architecture reference manual of the original processor. This is not trivial at system design.
2. It is not always safe to apply the CFD compiler transformation. Legality of a transformation is a must; thus, for CFD to be applied, the compiler needs to prove the safety of the transformation, which is not possible in many cases due to potential aliasing among the pointers/arrays in the loop, as well as loop-carried dependences.
3. The CFD transformations may prevent/impact other transformations by the compiler resulting in a net loss.

It is important to note that BOSS, on the other hand, works as a branch-hint, and thus resolves the above issues as we describe further below; to appreciate the difference of *CFD-transformation* vs *BOSS's hint-provisioning*, it may help to look at an analogy on data-access case: imagine a loop has long-latency misses on a 'load' instruction whose backslice is loop-separable; one can create a corresponding pre-execute loop to issue those 'load' instructions earlier and push the data to a newly introduced in-hardware queue so that they are then poped and consumed by the main loop; this is analogous to the CFD-transform. On the other hand, one can instead keep the original main loop intact, but put software-prefetch instructions in the pre-execute loop with same data-addresses as those 'load' instructions; this is analogous to BOSS's hint-provisioning. In the latter case, even if all the executed software-prefetch instructions are dropped by the processor (which is still a totally corret implementation of an ARM processor as per ARM ISA architecture reference manual), or a context-switch happens or a function is called in between software-prefetch and the target 'load' instruction, the program functionality of the latter approach is still correct because all those software-prefetch operations were *hopeful hints* or *assists*; of course the performance may be hindered in such cases, but it is tolerable if you note that the former approach totally breaks in these cases if the save/restore is not done.

Thus, compared to CFD,

(i) No library change or OS support is necessary by BOSS for program correctness. The branch-direction hints that it provides may even be totally ignored by a less capable processor core, or get lost upon a call or a context-switch; the functionality of the program remains intact.

(ii) It is always safe to apply BOSS; the functionality of the program always remains right. Obviously, the downside is that BOSS has to compute the branch-condition twice: once in the pre-execute loop and once more in the main loop. In this paper, we showed that despite this additional overhead, still reasonable gains can be obtained.

(iii) Since the original main loop remains intact, any prior compiler transformation is still applicable.

We showed that despite the overhead, BOSS is still beneficial. On top of that, note below additional advantages enabled by the hint-nature of BOSS that CFD by nature cannot do:

**a. Partial coverage of the iteration-space of the target loop**

In many cases, there are early-exists from the loop (e.g. `break` or `return` under a condition). Thus the *typical* trip-count of the loop may differ from maximal trip-count checked (e.g. by the loop-end-check in a for-loop). Profile-guided optimizations (PGO), or even simple manual profiling, can reveal that typical trip-count. In such case, the pre-execute loop can be run only up to that typical trip-count instead of covering the entire iteration-space of the main loop.

Taking this one step further, profiling can show which iterations of the loop more often experience a branch-misprediction by the in-HW predictors, and the pre-execute loop can cover only that subset.

Thus, partial coverage applies to loops in which there is an early exit or there are specific loop iteration ranges in which traditional branch predictors have a harder time guessing correctly for the target branch within the loop's body.

**Usecase demonstration:** Fig. 9 shows `kill_or_connect()` function of leela benchmarks of SPEC 2017, which demonstrates a use case of partial coverage's early exit. To demonstrate the case, we created different micro-benchmarks each covering a subset of the iteration-space; these are represented by n-to-m in the results in Fig. 10 where n and m represent the lower-bound and the upper-bound of the pre-execute loop in Fig. 9 respectively. The results in Fig. 10, including IPC, instruction-overhead, and finally cycle-count (speedup) are normalized to the full-coverage case where the pre-execute loop covers all the 4 iteration from 0 to 3.

```
void FastBoard::kill_or_connect(int color, int vertex,
int n, int m ) {
    /* The pre-execute loop */
    for (int k = n; k <= m; k++) { //n-m bound-control
        int ai = vertex + m_dirs[k];
        int libs = m_libs[m_parent[ai]];
        bool outcome = (libs <= 1);
        BOSS_write(BOSS_CHANNEL_0, k, outcome);
    }
    /* End of pre-execute loop. */
    /* Main loop: */
    bool retval = false;
    for (int k = 0; k < 4; k++) {
        int ai = vertex + m_dirs[k];
        int sq = get_square(ai);
        int libs = m_libs[m_parent[ai]];
        if ((libs <= 1 // <- Target Branch
            && sq == !color) ||
                (libs >= 3 && sq == color)) {
            retval = true;
            break;
        }
    }
    return retval;
}
```

Fig. 9. Partial coverage of the loop iteration-space (from n to m, instead of the full range which is 0 to 3), demonstrated on a case from Leela benchmark of SPEC 2017.

As Fig. 10 shows, partial coverage can make a big difference on overheads and gains, and is an additional angle to tune BOSS for the highest gain.

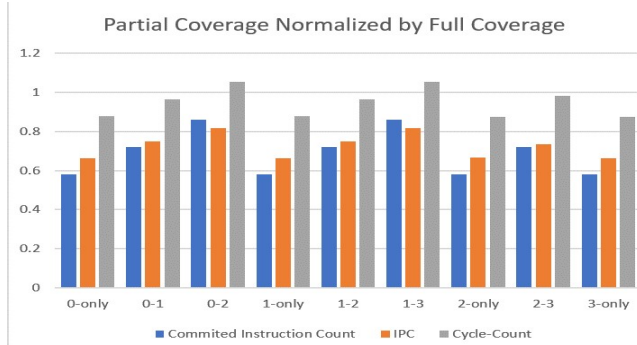

Fig.10. Various ranges of partial-coverage make a big difference in final gain obtained by BOSS.

  **b. Record-and-replay of branch-outcomes:** Since BOSS is a hint-passing mechanism, it can be used to replay a sequence of statically or dynamically recorded sequence of Taken/Not-taken branch outcomes that are known to be *mostly* (but not always) correct through profiling or analysis. Dynamic instances of the branch in the loop may show an irregular hard-to-predict pattern of Taken/Not-taken (e.g. 101100010 where each 1 shows a Taken and every 0 a Not-Taken instance of the branch), but then repeat (even partially) that pattern subsequently when the loop is observed again. In such case, BOSS can be employed without the pre-execute loop: Record-and-replay focuses on the idea that a branch within an inner loop (BIL) can repeat (fully or partially) its branching behavior across multiple outer loop iterations. In that case, the BOSS hint functionality can record BIL's behavior across one outer loop iteration and replay it for a subsequent outer loop iteration.

**Usecase demonstration:** This microbenchmark in Fig. 11 comes from leela's `save_critical_neighbours()` function with data values for arrays coming from original benchmark with ref.sgf as input. Through experiments, it can be seen that on average, 93% of a generation's target branch outcomes are repeated on subsequent generation, demonstrating an effective case for the record-and-replay functionality. Note how the imperfection of the matchings prevents a mechanism such as CFD to be used here.

```
void save_critical_neighbours(int color, int vertex,
movelist_t & moves, int & movecnt) {
    for (int k = 0; k < 4; k++) {
        int ai = vertex + m_dirs[k];
        int par = m_parent[ai];
        int lib = m_libs[par];
        bool cond = (lib <= 1);
        // record this generation's branch outcomes
        BOSS_write(BOSS_CHANNEL_1, k, cond);

        if (cond) {/* <- Target branch for replay in
                         the next generation */
            moves[movecnt++] = ai;
        }
    }
    return;
}
```

Fig. 11. Record-and-replay mechanism for partially-repeated patterns, demonstrated on a case from Leela benchmark of SPEC 2017.

Another use-case for record-and-replay is serverless functions in datacenters: the repeat-cycle of a branch may be so long that the micro-architecture context is lost when the branch is revisited. This is already observed in server workloads and serverless functions in datacenters [14], as well as in Android on mobile phones; such above save-restore mechanism as BOSS provides can be used here as well.

  **c. Correlation among different branches:** This case is based on the idea that a branch's behavior could tell us about the behavior of another branch within same or a different loop. The BOSS hint can make use of this correlation, and thus can help predict the behavior of this branch within the separate loop. Techniques such as Whisper [15] try to establish a correlation among branches, and then pass that correlation via a formula to the hardware to allow it predict the branch outcome. BOSS can do it in software for the same.

**Usecaes demonstration:** The microbenchmark in Fig. 12 shows the case that one branch's behavior within leela's `kill_or_connect()`, directly relates to the behavior of another branch in a separate loop within `kill_neighbours()` in Fig. 13. The root cause for this correlations is that both these functions have a common caller and are called closely next to each other by that caller; the differentiating factor for them is the `vertex` vs. `pos` values in the two functions, which are often equal as per experiments. The BOSS hint is able to take advantage of this relationship, and forward the (mostly) correct branching behavior to the first generation of target branch within `kill_neighbours()`. Yet again note how this correlation is statistical, but not mathematically provable; as a result, CFD cannot provide this functionality, but BOSS's hint nature does.

```
void FastBoard::kill_or_connect(int color, int vertex)
{
    bool retval = false;
    for (int k = 0; k < 4; k++) {
        ...
        int ai = vertex + m_dirs[k];
        int libs = m_libs[m_parent[ai]];
        // Targets a branch within kill_neighbours
        BOSS_write(BOSS_CHANNEL_2, k, (libs<=1));

        if ((libs <= 1 // <- Source branch
            && ...) {
            retval = true;
            break;
        }
    }
    return retval;
}
```

Fig. 12. BOSS allows to use the correlation among branches in different loops to resolve branches in a target loop. The designated source branch is correlated to the other branch in Fig. 13.

```
void FastBoard::kill_neighbours(…) {
   ...
   do {       ...
      for (int k = 0; k < 4; k++) {
         int ai = pos + m_dirs[k];

         if (m_square[ai] == kcolor) {

            int libs = m_libs[m_parent[ai]];
            if ((libs <= 1 // <- Target branch
                 && ... )
               ...
         }
      }
      pos = m_next[pos];
   } while (pos != vertex);
```

Fig. 13. Another delinquent branch in leela's kill_neighbours, whose first generation is correlated with the branches in kill_or_connect (Fig. 12)

## 5 SUMMARY AND CONCLUSION

BOSS can be best described as *branch counterpart of software data prefetching*. Accordingly, it by nature bears all the pros and cons of its data prefetching counterpart. BOSS is not a replacement for hardware predictors, in the same way that software prefetching is not a replacement for hardware prefetchers. Again similar to software prefetching, BOSS is a custom solution for certain niche cases where more general solutions do not perfectly fit, such as low-tripcount loops or for partial coverage of the loop iteration-space. Extending BOSS to also cover the loop-end branch is a trivial extension. We quantified the achievable gains by BOSS on a number of SPEC and other benchmarks, and explored the limits of its applicability.

**Discussion of limitations:** In the pre-execute mode, we limit BOSS usecases to non-nested branches to avoid having to insert branches in the pre-execute loop. If caller and callee functions both use BOSS at the same time (e.g. imagine the caller does the call in between the pre-execute loop and the main loop, or within the main loop), they should use different BOSS channels, or otherwise the callee would overwrite the branch-outcome values in the channel, and hence, the caller loses to use any remaining elements that used to be available in the channel. Obviously, save/restore is one solution, which is also relatively cheap and can be applied in the caller whenever calling a certain/potential BOSS-user function such as a library function. Saving of an already-open channel can also be done under the hood by the hardware when a new BOSS_open() is called on the same already-open channel by the callee; it can then be restored upon return from the callee. Note that each additional channel is quite inexpensive; each channel imposes only 83 bytes overhead (see §2.4). Finally, even if by mistake or otherwise same channel is used by the caller and callee (e.g. imagine the case of a call to a library function whose source is not available to the programmer who is willing to use BOSS), the program functionality remains intact because BOSS is only a hint and is safe to be wrong. Of course the performance may suffer in such erroneous case, and hence this should be avoided by proper save/restore, but the point is that even if not, the programs does not break.

Early-exits from the target loop is supported by the generation# mechanism (see §2.2). Skip (e.g. `continue` statement in C) is supported only after the target branch in the loop, because otherwise, the target branch would be a nested branch (which is not intended to be supported as said above) under the condition checked for that skip.

To reduce BOSS overhead in the pre-execute mode, we are evaluating selective runs of the pre-execute loop only when the target branch is heavily mis-predicted. PGO-based selective ranges of the iteration-space is another method for the same goal. As also mentioned in §4.2, record-and-replay and correlation-based usecases are other promising avenues that the additional overhead would be much less since they do not need a redundant compute of the branch condition.